\documentclass[iop,apj,twocolumn]{emulateapj}

\usepackage{color}
\usepackage{graphicx}
\usepackage{bbold}
\usepackage{ulem}
\usepackage{times}
\usepackage{natbib}          
\shorttitle{Accretion of Jupiter-mass Planets}
\shortauthors{Szul\'agyi et al.}

\begin{document}

\slugcomment{December 2013, accepted for publication in ApJ}


\title{Accretion of Jupiter-mass Planets in the Limit of Vanishing Viscosity}


\author{J. Szul\'agyi\altaffilmark{1}, A. Morbidelli\altaffilmark{1}, A. Crida\altaffilmark{1}, F. Masset\altaffilmark{2}}
\affil{$^{1}$University of Nice-Sophia Antipolis, CNRS, Observatoire de la C\^ote d'Azur, Laboratoire Lagrange, France}
\affil{$^{2}$Instituto de Ciencias F\'\i sicas, Universidad Nacional Aut\'onoma de M\'exico, PO Box 48-3, 62251 Cuernavaca, Mor., Mexico}
\email{jszulagyi@oca.eu}

\begin{abstract}
In the core-accretion model the nominal runaway gas-accretion phase brings most planets to multiple Jupiter masses. However, known giant planets are predominantly Jupiter-mass bodies.

Obtaining longer timescales for gas accretion may require using realistic equations of states, or accounting for the dynamics of the circumplanetary disk (CPD) in low-viscosity regime, or both. Here we explore the second way using global, three-dimensional isothermal hydrodynamical simulations with 8 levels of nested grids around the planet.

In our simulations the vertical inflow from the circumstellar disk (CSD) to the CPD determines the shape of the CPD and its accretion rate. Even without prescribed viscosity Jupiter's mass-doubling time is $\sim 10^4$ years, assuming the planet at 5.2 AU and a Minimum Mass Solar Nebula. However, we show that this high accretion rate is due to resolution-dependent numerical viscosity.

Furthermore, we consider the scenario of a layered CSD, viscous only in its surface layer, and an inviscid CPD. We identify two planet-accretion mechanisms that are independent of the viscosity in the CPD: (i) the polar inflow -- defined as a part of the vertical inflow with a centrifugal radius smaller than 2 Jupiter-radii and (ii) the torque exerted by the star on the CPD. In the limit of zero effective viscosity, these two mechanisms would produce an accretion rate 40 times smaller than in the simulation.
\end{abstract}


\section{Introduction}

How exactly giant planets form is still one of the most puzzling questions in today's planetary science with lots of dark patches in the picture. The most popular giant planet formation theory is the core accretion model by \citet{BP86}.  There are three main stages of formation in this model. First, a planetary core is formed and starts attracting the gas within its Bondi-radius. When the core reaches $12-16 \rm{M_{\oplus}}$ the gas envelope starts to contract quasi-statically while the accretion rates increases \citep{Pollack96}. This second stage takes the longest time, of the order of a few million years. The final, runaway gas accretion phase starts when the envelope and core masses are approximately equal. This phase should not stop until the planet has opened a deep gap in the gas disk. However, this happens only when the planet reaches a mass of 5-10 Jupiter-masses \citep{Kley99,LDA06}. Thus Jovian-mass planets can double their mass on an order of $10^4$ years. 

Thus, one should expect to observe a dichotomy in the mass distribution of planets\,; planets should be either smaller than $\sim30$ Earth masses, i.e. those that did not reach the phase of runaway gas accretion, or larger than a few Jupiter-masses, i.e. those that entered and completed the fast runaway growth phase. Planets in between these two mass categories should be extremely rare. This is the converse of what is observed (e.g. \citealt{Mayor11}). Thus, there is a need to understand what sets the final mass of a giant planet.

An obvious possibility to stop accretion is that the gas disappears while the planet is still growing. However, the lifetime of gaseous proto-planetary disks is of the order of a few million years \citep{Haisch}, which is much longer than the runaway growth timescale ($10^4$yr). It is very unlikely that the disappearance of the disk can happen at the right time to stop the runaway growth of the planet. Another possibility is that a planet cannot accrete more gas than what is delivered to its orbit by viscous accretion, i.e. it cannot grow faster than the star accretion rate. In general, the accretion rate observed in proto-planetary disks is on the order of $10^{-8}-10^{-7} \rm{M_{\odot}}/\rm{yr}$. This would allow the accretion of Jupiter's atmosphere in $10^4$--$10^5$~yr which, is too short relative to the lifetime of the disk. If one requires that Jupiter takes a million years to accrete its envelope, then its runaway growth needs to be limited by a stellar accretion rate of $10^{-9} \rm{M_{\odot}}/\rm{yr}$. But at this very low rate the disk photoevaporates rapidly (i.e. a few $10^5$ years, see e.g. \citealt{Koepferl13,Gorti09,Szulagyi12}). Thus a Jupiter-mass of gas is unlikely to be accreted by the planet. A very accurate tuning between the viscous accretion rate, the photoevaporation rate, and the runaway growth seems to be needed to allow a planet to grow to Jupiter-mass but not beyond this limit. Something must be still missing from the picture.

What we need is a mechanism that slows down runaway growth. So it occurs on a timescale comparable to the disk's lifetime. A possibility is that the circumplanetary disk acts as a regulator of gas accretion rate onto the planet. Before the gas is accreted by the planet, it has to pass through the circumplanetary disk (CPD) because of angular momentum conservation. The actual accretion rate of the planet then depends on the timescale for angular momentum transport within the CPD. If the circumplanetary disk has a very low viscosity, then the transport of angular momentum through this disk is inefficient and gas accretes onto the planet at a slow rate. In this situation the observed mass spectrum of the giant planets is set by the competition between gas accretion and gas dissipation \citep{Rivier12}.

There are good reasons to think that the viscosity is very low in the CPD  (although see \citealt{Gressel13} for an opposite view). Firstly, the planets are thought to be formed in a dead zone of the circumstellar disk, where the viscosity is low \citep*{Thommes08,ML11}. Secondly, because the CPD is shadowed by the circumstellar disk and by the remaining gas in the gap, its irradiation geometry may be unfavorable for ionization \citep{Turner10,Turner13}. Finally, the large orbital frequencies in the CPD make the magnetic Reynolds number too small to derive the magneto-rotational instability \citep{Fujii11,Turner13,Fujii13}.

Motivated by these considerations, in this paper we study the dynamics of the CPD in detail. 

Our simulations are similar to those in \citet{Machida10} and \citet{Tanigawa12} with one main difference. Instead of using a local shearing sheet approximation, we perform global disk simulations. This better allows us to study the connection between the circumstellar disk and the circumplanetary disk i.e. opening of a gap, gas flow through the gap etc. Moreover, we investigate in more detail the accretion rate of a Jovian-mass planet in the limit of vanishing viscosity.  To do this, it is not enough to perform simulations with no prescribed viscosity as in \citet{Tanigawa12}, because every numerical simulation is affected by numerical viscosity. We need to identify the various accretion mechanisms and distinguish between those dependent on viscosity and those independent of viscosity (i.e. polar inflow from the circumstellar disk, loss of angular momentum due to shocks, stellar torque exerted on the CPD, etc.) and evaluate their magnitude. 

A well-known crucial issue for simulating gas accretion onto a planet is the choice the equation-of-state (EOS). Several works have stressed the need to use an adiabatic EOS -- possibly complemented by a recipe for radiative cooling -- in order to study planet accretion (\citealt{DAngelo03,KK06,PM08,AB09}). However, the differences with the isothermal EOS are fundamental for small-mass planets (up to Saturn's mass), but less for Jupiter-mass planets. In the latter case the flow of gas is mostly dominated by the planet's gravity. CPDs definitely form around Jupiter-mass planets and the differences between isothermal and radiative simulations are mostly limited to the mass of the CPD and its scale height \citep{DAngelo03,AB09,Gressel13}. Thus, we prefer to use the isothermal EOS, with temperature dependence on stellar distance (hereafter  locally isothermal), for multiple reasons. The first is that we wish to focus our paper on the role of numerical viscosity and on viscosity-independent transport mechanisms within the CPD, which have never been thoroughly discussed before; these considerations should be independent of the EOS assumed. Second, radiative simulations imply additional, badly constrained parameters such as those in the prescription for the opacity laws (e.g. \citealt{AB09,Bitsch}). We want to focus  the discussion on the objectives stated above without distraction. Third we wish to make direct comparisons particularly concerning the differences between our global disk simulations and shearing-sheet studies \citep{Machida10,Tanigawa12}, and the latter have been done with isothermal EOS. Finally, this paper is the first in a series of future studies, therefore we wish to begin with the most simple case and build on it incrementally. Nevertheless, for each result that we present, we will state to what extent we expect it to be valid or different in a non-isothermal context.

The paper is structured as follows. In Section \ref{setup}, we describe the setup of our hydrodynamic simulations. This is followed by the results on the structure of the CPD in Section \ref{structure}. Then, Section \ref{accrete} discusses our findings on the accretion mechanism. Section \ref{discussion} reports discussions and perspectives. Finally, Section \ref{conclusion} summarizes the conclusions of our work.

\section{Setup of the Simulations}
\label{setup}

\subsection{Physical Model}

We performed hydrodynamic simulations of an embedded Jupiter-mass planet in a circumstellar disk. The coordinate system was spherical and centered on the star. The planet was on a fixed circular orbit. The units of the code were the following\,: the unit mass was the mass of the star ($M_*$), the length unit was the radius of the planetary orbit ($a$), and the unit of time was the planet's orbital period divided by $2\pi$. Consequently, the gravitational constant ($G$), the planet's angular momentum and orbital (angular) velocity $\Omega=\sqrt{GM_*/a^3}$ are unity. The frame was co-rotating with the planet. Our planet was placed at coordinates\,: 0, 1, $\frac{\pi}{2}$ (azimuth, radius, co-latitude, respectively). In our simulation we used an azimuth range of $-\pi < \theta <\pi$, a radius range of 0.41 $< r/a <$ 2.49, and a co-latitude range of 3 times the pressure scaleheight\,: [$1.42 < \phi < \frac{\pi}{2}$]. We assumed symmetry relative to the midplane, therefore only half of the circumstellar disk was simulated.

The initial surface density is $\Sigma=\Sigma_0(r/a)^{-1.5}$ where $\Sigma_0=6 \times 10^{-4}$ (in code units). Note that, because our equation of state is locally isothermal (see below), the equations are linear with $\Sigma_0$ except for the indirect term. Since our $\Sigma_0$ is small, the indirect term is negligible, so that our results scale linearly with $\Sigma_0$. We chose $\Sigma_0$ such that with $M_*=M_\odot$ (the solar mass) and $a=5.2$~AU, our initial surface density profile is very close to \citet{Hayashi}'s Minimum Mass Solar Nebula (MMSN) our mass unit to be the solar mass $M_\odot$ and our length unit to be $a_{jup}=5.2$ AU, in order to set our initial surface density profile very close to \citet{Hayashi}'s Minimum Mass Solar Nebula (MMSN). For the reader's convenience, we scale our results with $\Sigma_0/\Sigma_{MMSN} \sqrt{a/a_{jup}} (M_{\odot}/M_*)$, this way one can easily compare his/her results with ours. Because MMSN is proportional to $r^{-3/2}$, and our $\Sigma_0$ dimension is $M_*/a^2$, the general relationship between $\Sigma_0$ and $\Sigma_{MMSN}$ is\,: $\Sigma_0=\Sigma_{MMSN} \sqrt{a_{jup}/a} (M_*/M_{\odot})$, therefore, we are using this scaling in the followings.

In our ``nominal'' simulation the gas was set to be inviscid, i.e. there is no prescribed viscosity in the fluid equations. We stress, however, that fluid is nevertheless affected by numerical viscosity, whose effects will be quantified by changing the resolution of the numerical grids (see below). For comparison purposes, we also ran a simulation with an $\alpha$-prescribed viscosity \citep{sunayev} adopting $\alpha = 0.004$. Hereafter, we will refer to this as our ``viscous simulation". Notice, that $\alpha$ sets a viscosity which is a function of heliocentric distance (radius). However, since the CPD size is small, the viscosity in the CPD can be considered uniform.

As discussed in the introduction, the equation-of-state (hereafter, EOS) is locally isothermal\,: $p=c_s^2\rho$ with disk aspect-ratio $H/r=0.05$, where $H=c_s/\Omega$ (here, $c_s$ is the speed of sound, $\Omega$ indicates the angular velocity, $p$ stands for the pressure and $\rho$ is the volume density). No magnetic field was included in the computations. 

The planetary mass in the simulations was set to $10^{-3}$ stellar masses, in order to study planet accretion at a Jupiter-mass. 
However, we did not introduce the planet with its full mass from the beginning. Instead, we prescribed a smooth mass growth of the planet as $\sin(t/t_0)$, where $t_0$ was 5 planetary orbital periods. This was done for numerical reasons, so that the gas had the time to adapt to the presence of a progressively more massive planet. The simulations overall have been ran for 238 planetary orbits.

\subsection{Numerical Model}

For the simulations, we used a three-dimensional nested-grid code, called {\sc Jupiter}. The {\sc Jupiter} code solves the Riemann-problem at every cell boundary \citep{Toro_book} to ensure the conservation of mass and the of three components of momentum:
\begin{equation}
\rho_t+\nabla \cdot (p\mathbb{v})=0
\end{equation}
\begin{equation}
(\rho v)_t+\nabla \cdot (\rho \mathbb{v} \otimes \mathbb{v} + p \mathbb{I})=0
\end{equation}
where $\rho$ is the density, $p$ the pressure, $\mathbb{v}$ the velocity vector, and $\mathbb{I}$ indicates the identity matrix. The use of a Riemann-solver makes {\sc Jupiter} particularly suited to treat shocks, contrary to the van Leer method \citep{Leer77}. The Riemann-solvers implemented in the {\sc Jupiter} code are approximated solvers based on the exact solution\,: Two-Shock solver, and Two-Rarefaction solver \citep{dVB}.

The timestep in the simulation is adapted by the code, in order to satisfy the Courant-Friedrichs-Lewy condition (CFL condition) for all levels of mesh resolution:
\begin{equation}
C=\Delta t \sum_{i=1}^{3} \frac{v_{x_{i}}}{\Delta x_i} \leq 1.0
\end{equation}
where $C$ is the Courant number, $i$ represents the number of dimensions, $x_i$ means the spatial variables, and $v$ indicates the velocity. The timestep at a given level can be the same as the timestep on the higher resolution level, or it can be twice that timestep, in which case two iterations are performed on the finer level while one iteration is done on the coarser level. This latter technique is called timestep subcycling. We use an adaptive subcycling procedure, which will be described in a forthcoming publication, in order to obtain the maximum speed up of the code (the highest possible ratio of physical time over wall clock time).

The full viscous stress tensor is implemented in the code in three geometries: Cartesian, cylindrical and spherical. The spherical implementation, that we use in this work, has been tested thoroughly in a prior work \citep{FLM11}.

We employed a system of 8 nested grids, where at level 0 (i.e. in the coarsest grid) the resolution was $628\times208\times15$ cells for the directions of azimuth, radius, co-latitude, respectively.
Each additional grid was added after the gas reached a stationary configuration and each of them was centered on the planet. The size of the cells in a grid at a given level was $\frac{1}{2}$ in each spatial direction of the cell size of the next larger grid. Table~\ref{levels} contains the number of cells on each level and the grid boundaries. In the finest level, the cell length was $7.82 \times 10^{-5}a$, which is 0.113\% of the Hill-radius of the planet, and 87\% of the radius of the present day Jupiter, assuming the planet orbits 5 AU away from the star. The cells had the same length in every directions (i.e. they were cubes), and the radial spacing in between them was arithmetic.

\begin{table*}
\centering
  \caption{Number of cells on different grid levels}
 \label{levels}
\scriptsize
  \begin{tabular}{p{.7cm}p{1.2cm}p{1cm}p{1.2cm}p{3.7cm}p{3.5cm}p{2.5cm}}
  \hline
   Level    &  $N^{\circ}$ of cells in azimuth  &  $N^{\circ}$ of cells in radius &  $N^{\circ}$ of cells in co-latitude & Boundaries of the levels in azimuth [rad] &  Boundaries of the levels in radius [a] & Boundaries of the levels in co-latitude [rad]\\
 \hline
\hline
       0	&	628	&	208	&	15 & [$-\pi$, $\pi$] & [0.41, 2.49] & [1.42, $\pi/2$]\\
       1	&	112	& 	112	& 	24 & [-0.27735, 0.27735] & [0.72264, 1.27735] & [1.451041, $\pi/2$]\\
       2	&	112	&	112	&	40 & [-0.138675, 0.138675] & [0.861325, 1.138675] & [1.47137, $\pi/2$]\\
       3	&	112	& 	110	&	56 & [-0.0693375, 0.0693375] & [0.9306625, 1.0693375] & [1.5014588, $\pi/2$]\\
       4	&	112	&	110	& 	56 & [-0.03466875, 0.03466875]& [0.96533125,1.03468875] & [1.5361276, $\pi/2$]\\
       5	&	112	&	112	& 	56 & [-0.017334375, 0.017334375] & [0.98266562, 1.017334375] & [1.553462, $\pi/2$]\\
       6	&	112	&	112	& 	56 & [-0.0086671875, 0.00866719] & [0.99133281, 1.0086671875] & [1.5621291, $\pi/2$]\\
       7	&	112	& 	110	& 	56 & [-0.0043335938, 0.00433359] & [0.99566641, 1.0043335938] & [1.566462737, $\pi/2$]\\
\hline
\end{tabular}
\end{table*}

On level 0 we used reflecting boundary conditions except in the azimuthal direction where we used periodic conditions. The communication between the grids at level $i$ and $i+1$ (where $i=0,\ldots,7$) were done through ghost cells with multi-linear interpolation.

To test the effects of numerical viscosity, we also ran a simulation with a twice finer resolution ($1256\times416\times30$), that we call hereafter the ``high-resolution simulation". Because the simulation is extremely slow at this resolution, we did not start it from time zero, but from the output of the nominal resolution at 238 orbits; we re-binned the gas on the new grids and then ran the code for an additional 10 orbits.

The planet was not modeled, but was treated as a point-mass placed in the corner of four cells on the midplane. In order to avoid a singularity, the planetary potential was smoothed as\,:
\begin{eqnarray}
U_p&=&-\frac{G M_p}{\sqrt{x_d^2+y_d^2+z_d^2+{r_s}^2}}
\end{eqnarray}
where $x_d=x-x_p$, $y_d=y-y_p$, and $z_d=z-z_p$ are the distance-vector components from the planet in Cartesian coordinates. The smoothing length $r_s$ was set equal to the cell size in levels 0--4. From level 5 on we used 2 cell sizes. Moreover, when introducing levels 6 and 7, we progressively decreased $r_s$ from the value used in the previous level to its desired final value. For example, when introducing level 7, first the same smoothing length was applied as on level 6 (which is equal to 4 cell sizes on level 7), but then it was decreased in time with a sinusoidal function until reached $r_s=2$ cells. This technique was done to allow the gas to adapt to a gradual change of the gravitational potential.

Because of the isothermal character of our fluid equations, a huge amount of mass tends to pile up in the few cells neighboring the point-mass planet. This causes numerical instability. Thus, we applied a density cut\,: if the volume density reached $1.42 \times 10^5$ (in the code's units), then the volume density of that cell was limited to this value (hereafter we refer to this as the ``mass-cut"). We keep track of the mass removed in this operation, from which we compute the planet accretion rate. However, the mass of the planet that enters in the gravitational potential was not changed. This is because we are interested in the accretion rate of a Jupiter-mass planet, and not in the growth of the planet itself. 

\section{Structure of the circumplanetary disk}
\label{structure}

In this section we describe our results about the circumplanetary disk structure: its vertical structure, its radial structure and the radial flow in the midplane. All the results that we present are from our nominal simulation unless we specify otherwise.

\begin{figure}
\includegraphics[width=\columnwidth,clip=true]{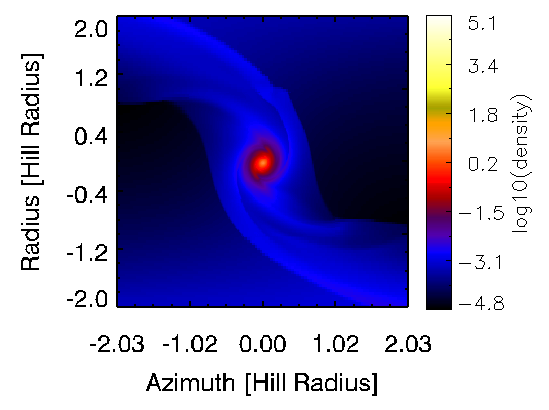} 
\caption{Volume density map of our inviscid, low-resolution simulation on the midplane using data from levels 2-7. The planet is in the middle of the figure. The planet clearly opened a gap and  the spiral density wave launched by the planet connects the circumstellar disk with the CPD. Here and in the following figures, with "Azimuth" and "Radius" we mean the distance from the planet in the azimuthal and radial direction.}
\label{gap}
\end{figure}

Because the grid at level 0 covers the circumstellar disk globally, the planet can open a gap around its orbit (see Figure \ref{gap}). This was not the case in the simulations of \citet{Machida10}, and \citet{Tanigawa12} because of the shearing-sheet approximation they adopted. Fig. \ref{gap} also shows that the density wave launched by the planet in the circumstellar disk smoothly joins the CPD and spirals into it down to the planet (see also Fig. \ref{streamline}).

\subsection{The vertical structure of the circumplanetary disk}
\label{vertical}

We start by discussing the vertical structure of the CPD.
For this purpose, it is convenient to characterize the CPD based on the z-component (in Cartesian coordinates) of the specific angular momentum in respect to the planet, normalized to the Keplerian value\,:

\begin{equation} \label{eq:angmom}
L_z=\frac{x_dv_y-y_dv_x+(x_d^2+y_d^2) \Omega}{\sqrt{G M_{P} \sqrt{x_d^2+y_d^2}}}
\end{equation}
where $v_x$, $v_y$ are the velocity-components transformed to Cartesian coordinates in the co-rotating frame.

\begin{figure}
\includegraphics[width=\columnwidth,clip=true]{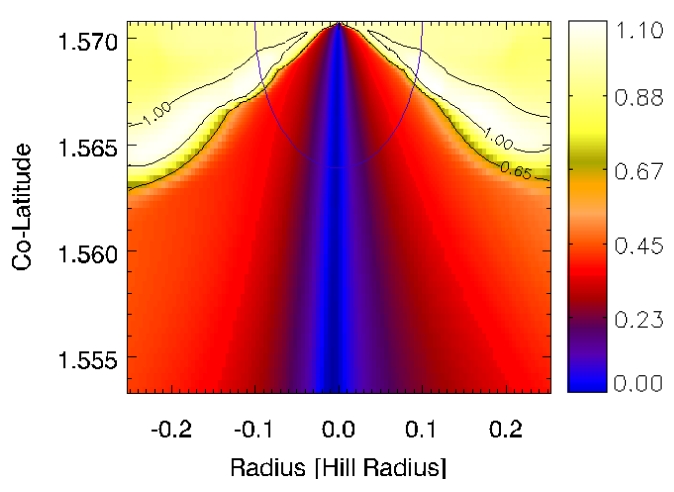} 
\caption{A vertical slice of disk passing through the planet's location, showing in colors the values of the normalized specific angular momentum $L_z$ of our inviscid, low-resolution simulation. A value of $L_z \,\approx\, 0.65$ separates the CPD (see the corresponding contour line) from the environment. One can see that the gas near the midplane is sub-Keplerian (yellow), while on the surface layer of the disk it is slightly super-Keplerian (white region bounded by the contour line 1.0). The blue-violet colors correspond to gas that is falling almost vertically towards the CPD. The blue circle around the planet symbolizes the $\frac{1}{10}$ of the Hill-radius.}
\label{angmom}
\end{figure}

\begin{figure}
\includegraphics[width=\columnwidth]{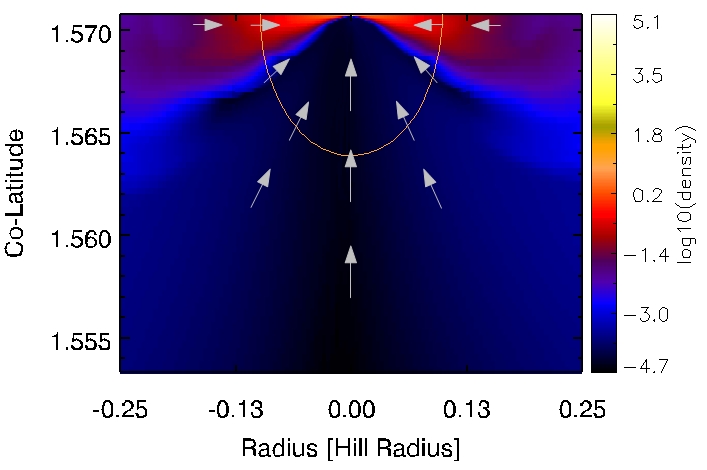} 
\caption{The same as Fig.~\ref{angmom}, but showing in colors the volume density of the gas of our nominal simulation; a few velocity vectors schematize the directions of the flow. Notice the vertical inflow, as well as the accreting flow in the CPD midplane. Again, the orange circle around the planet symbolizes the $\frac{1}{10}$ of the Hill-radius.}
\label{densvelo}
\end{figure}

Fig. \ref{angmom} represents a vertical slice at azimuth = 0.0 of the $L_z$ distribution in the neighborhood of the planet, which is located at the center of the upper axis. We see that $L_z$ rapidly drops from $\sim 1$ to $\sim 0.5$ at a location where the density shows a clear discontinuity (see Fig. \ref{densvelo}). Therefore, hereafter we define the CPD as the region where $L_z$ is larger than 0.65 (see the corresponding contour line in Fig. \ref{angmom}). Even this value is arbitrary but, given the steep gradient of $L_z$ near the surface of the disk, changing this threshold would not change significantly the results presented below.

The Fig. \ref{angmom} shows that the gas in the midplane and near the midplane is sub-Keplerian (similarly to \citealt{Tanigawa12,Uribe13}). However, notice that near the upper layer of the disk the flow is slightly super-Keplerian, i.e. in the region bounded by the contour line $L_z=1$. This is due to the fact that the disk is very flared, so that the radial pressure gradient near its surface is positive. 
In the viscous simulation, however, this super-Keplerian near-surface layer does not exist. This is due to the higher viscosity, that limits the vertical shear in the CPD.

The gas located below the CPD is falling towards the CPD with a large vertical velocity, as indicated by the arrows on Fig.~\ref{densvelo} (see also \citealt{AB09}, and \citealt{Tanigawa12}). As pointed out in \citet{Tanigawa12} the sharp vertical boundary of the CPD clearly visible in the $L_z$ and density maps is due to a shock front generated by the vertical influx. As in \citet{Tanigawa12}, we also notice from Fig.~\ref{angmom} that the vertical inflow hits the CPD with a value of $L_z$ that is much smaller than that characterizing the CPD at the same location. Thus, the vertical inflow slows down the rotation of the CPD, promoting  radial infall at the surface of the CPD.

\begin{figure}
\includegraphics[width=\columnwidth]{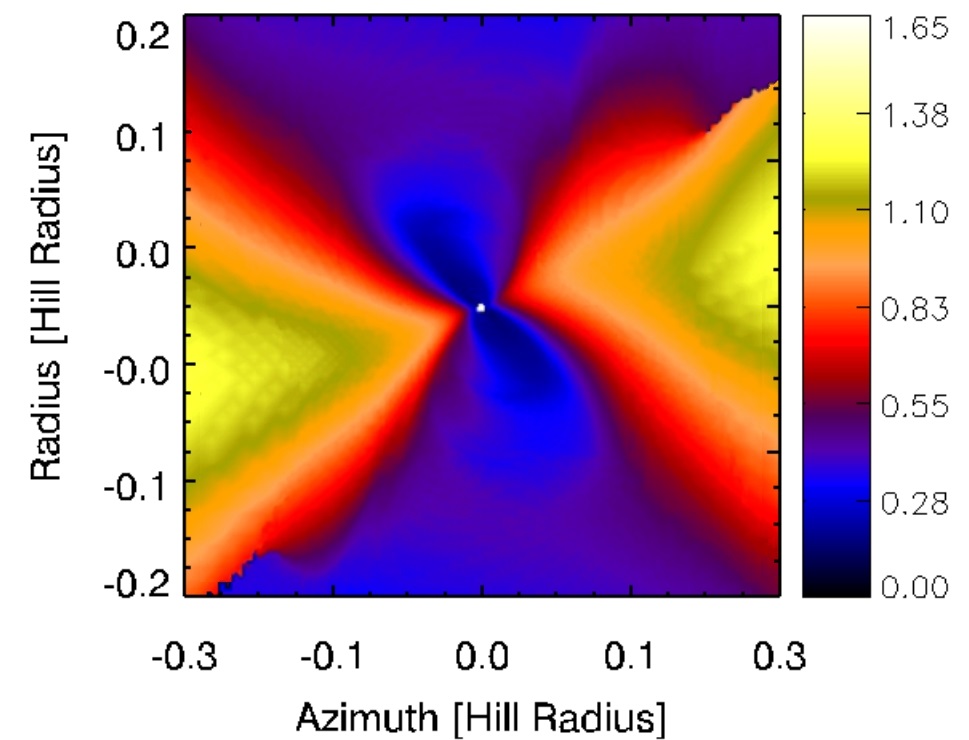} 
\caption{A color map showing the aspect ratio $z_{CPD}/d$ as a function of azimuth and radius of our inviscid, low-resolution simulation. The planet is placed at the center of the plot. At any given distance from the planet, the aspect ratio changes considerably with the planetocentric azimuths. Thus, the CPD has a ``wavy" surface structure. }
\label{wavy}
\end{figure}

\begin{figure}
\includegraphics[width=\columnwidth]{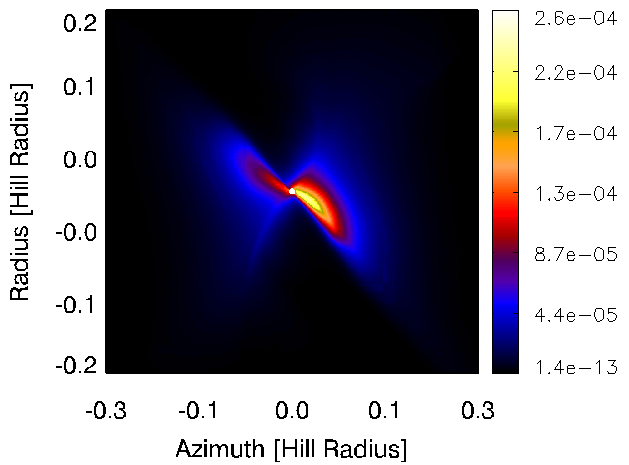} 
\caption{The same as Fig.~\ref{wavy}, but showing in colors the value of $\rho v_z^2$, representing the ram pressure exerted by the polar inflow on the CPD surface. It can be seen that the pressure of the inflow is higher along a diagonal  line oriented from top-left to bottom-right. Thus, the CPD is compressed along this line and has a minimum aspect ratio (see Fig.~\ref{wavy}) along a line rotated by $\theta_{delay}$ relative to the highest pressure line (see text).}
\label{influx}
\end{figure}

We find that the vertical influx has also a strong influence on the aspect ratio of the CPD. First of all, we remind that the pressure scale height of the CPD at hydrostatic equilibrium is $H_{CPD}\equiv c_s/\omega$, where $\omega=\sqrt{GM_p/d^3}$ is the angular velocity around the planet, and $d=\sqrt{x_d^2+y_d^2}$ indicates the distance from the planet. The sound speed ($c_s=0.05\,r^{-1/2}$)  is almost constant in the CPD in our locally isothermal simulation. So, we expect the aspect ratio of the CPD to be $H_{CPD}/d=\frac{c_s}{\sqrt{GM_p}}\,d^{1/2} \,\approx\, 1.6(d/a)^{1/2}=\, 0.16\,(d/0.01a)^{1/2}$, which is very thick and flared.
As we show in Fig.~\ref{wavy}, the surface of the CPD defined by $z_{CPD}$ (i.e. the uppermost z-ccordinate where $L_z\approx0.65$) is indeed strongly flared, but its aspect ratio has also a strong dependence on the azimuth relative to the planet. In fact, $z_{CPD}/d$ is changing from $\sim20\%$ to $> 100\%$. To our knowledge, this wavy surface has not been described yet in the literature. The wavy surface pattern is due to the dynamical pressure of the vertical mass inflow, which is not uniform in planetocentric azimuth (see Fig. \ref{influx}); it is maximal along an axis close to the axis of the spiral arm.

Now, let's take a Lagrangian approach and consider fluid elements in the CPD orbiting on circles centered on the planet; because the pressure due to the vertical inflow has two maxima they will feel a maximum of pressure twice per orbit. But it takes a time $t_{delay}\approx H_{CPD}/c_s$ for the gas in the CPD to react to the pressure pulse. In this time the fluid elements rotate by an angle $\theta_{delay}=\omega \times t_{delay}$, which is the angle between the axis marking the minimum height of the CPD and that marking the maximum pressure. Because  $\omega \times t_{delay}=\omega/H_{CPD}/c_s=1$, this angle is independent of the distance from the planet $d$. The comparison of the Figs. \ref{wavy} and \ref{influx} clearly shows an angle of order unity (in radians) between the maximum pressure and the minimum CPD height. A toy-model is presented in the Appendix~\ref{appendix} about how the pressure of the vertical influx leads to the observed structure of the CPD.

\begin{figure}
\includegraphics[width=\columnwidth]{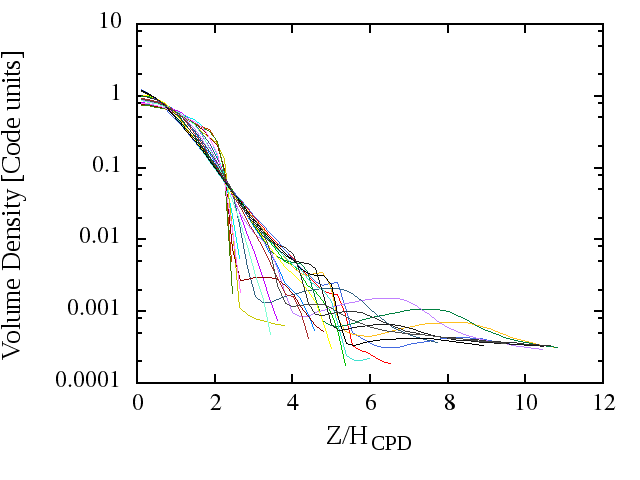} 
\caption{The volume density versus the z coordinate at $d \approx 0.058$ Hill-radii, for various azimuths relative to the planet. The vertical structure of the CPD changes with azimuth with fixed knots as a stationary wave. The data are from our inviscid, low-resolution simulation.}
\label{disk_surface}
\end{figure}

Fig. \ref{disk_surface} shows the vertical density distribution in the CPD from $z=0$ to $z=z_{CPD}$ at a given radius for various values of the azimuth. The mass is conserved along an orbital period, so the integral of each density curve is the same. On top of the expected equilibrium Gaussian shape, one can notice oscillations with two knot points where the density does not change with azimuth. This is reminiscent of stationary waves.

Simulations implementing an adiabatic EOS (e.g. \citealt{AB09, AB12,Gressel13}) also find that the vertical inflow is the main feeding mechanism for the CPD. The CPD however has a larger scale height as it is hotter and the boundary between the disk and the vertical flow is less sharp than in our isothermal simulations. We will come back to this last, important issue in Sect. \ref{discussion}.

\subsection{Radial structure of the circumplanetary disk}

We now move to discuss the orbital motion inside the CPD and the location of its outer radial boundary.

The specific normalized angular momentum $L_z$ on the midplane declines with the distance from the planet \citep{AB09, Tanigawa12}, but it does not show a steep gradient like the one in the vertical direction near the surface of the disk. Therefore, previous authors assumed arbitrary limits in $L_z$, obtaining different radial extensions for the CPD. 

We think that it is more meaningful to look at the orbital motion of the fluid elements, defining the CPD as the region where the orbits are basically circular. Quasi-circular orbits may have a small value of $L_z$ if the CPD is strongly sub-Keplerian due to a steep radial pressure gradient, but they are clearly part of a disk.

\begin{figure}
\includegraphics[width=\columnwidth]{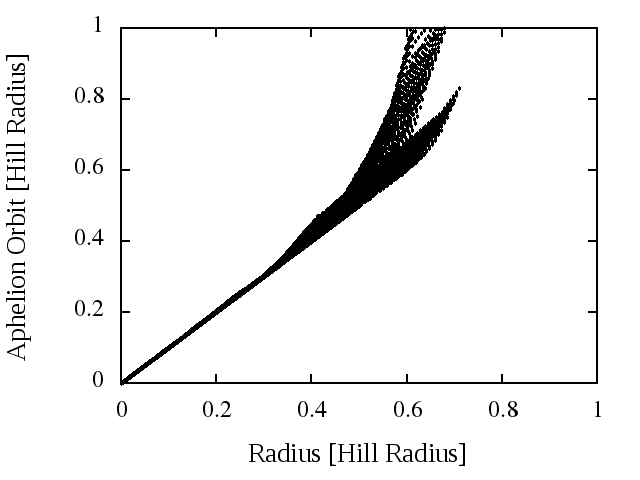} 
\caption{Orbital apocenter as a function of planetocentric radius for the cells on the midplane in the vicinity of the planet. As long as the points lie on a line of slope 1, the streamlines in the disk are circular. One can read from the figure that the CPD of our inviscid, low-resolution simulation is quite circular up until $\sim0.48$ Hill radii.}
\label{ecc}
\end{figure}

In order to visualize easily where the orbits of the disk are quasi-circular, we proceeded as follows. First, we calculated the semi-major axis $a$ and the eccentricity $e$ in every cell from the cell's coordinates, the recorded velocities and the planetary potential; then we plotted the apocenter of the orbit $Q=a (1+e)$ versus the planetocentric radius $d$ of the cell. If, at a given radius every cell, whatever its planetocentric azimuth, appears to be at apocenter ($Q=d$), that obviously means the streamline in the disk is circular, although sub-Keplerian. We can see on Fig. \ref{ecc} that this is the case up to $\sim0.48$ Hill radii. If we use this definition for the radial extent of the CPD, then the disk is a bit wider than the previously recorded radial extensions of $\sim0.1-0.3$ Hill radii \citep{Tanigawa12,AB09}.

We remark however that the eccentricity of the streamlines in the disk depends on the viscosity. In fact, as we will see in Sect.~\ref{midflow}, the streamlines are eccentric if they are shocked at the passage through the wave generated by the stellar tide. The smaller is the effective viscosity -- prescribed or numeric -- in the CPD, the closer to the planet the wave propagates and shocks. Thus, defining the CPD as the region where streamlines are circular may lead to the uncomfortable situation that the disk may become vanishingly small in the ideal limit of zero viscosity. In fact, in our viscous simulation, circular orbits extend up to $\sim0.55$ Hill radii and in the high resolution simulation, which halves the numerical viscosity, they extend only up to $\sim0.28$ Hill radii. This is an important point that should be kept in mind when analyzing the results of simulations, regardless if conducted with an isothermal or adiabatic EOS.

In alternative, we may define the radial extent of the CPD as the largest circle from which streamlines wrap around the planet at least once before becoming unbound, in either the forward or backward integration. If we adopt this definition, the radius of our CPD is approximately 1/2 to 3/4 of the Hill radius.

\begin{figure}
\includegraphics[width=\columnwidth]{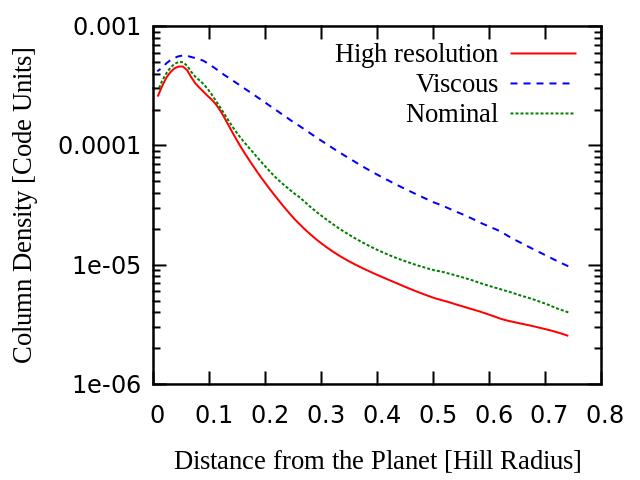} 
\caption{The column density profile of the CPD in respect to the distance from the planet. Each curve refers to a different simulation, as labeled.}
\label{coldens}
\end{figure}

On Fig. \ref{coldens} our CPD's column density ($\int{\rho dz}$) profile can be seen. We have less massive CPD than \citet{Tanigawa12}, probably because our global disk simulation contained a planetary gap in contrary to the sheering sheet box simulations. Instead the column density at 0.1 Hill radius in our CPD ($\sim 100$g/cm$^2$ for Jupiter at 5 AU in a MMSN) is comparable to that in the radiative simulations with reduced opacity of \citep{AB09} and with the most viscous simulation in \citet{DAngelo03}. 

\subsection{Flow in the midplane of the circumplanetary disk} 
\label{midflow}

There is a debate in the literature about the direction of the radial flow on the midplane of the CPD. \citet{AB09} found inflow in their simulations, while \citet{Tanigawa12,KK06,AB12} found outflow. 

\begin{figure}
\includegraphics[width=\columnwidth]{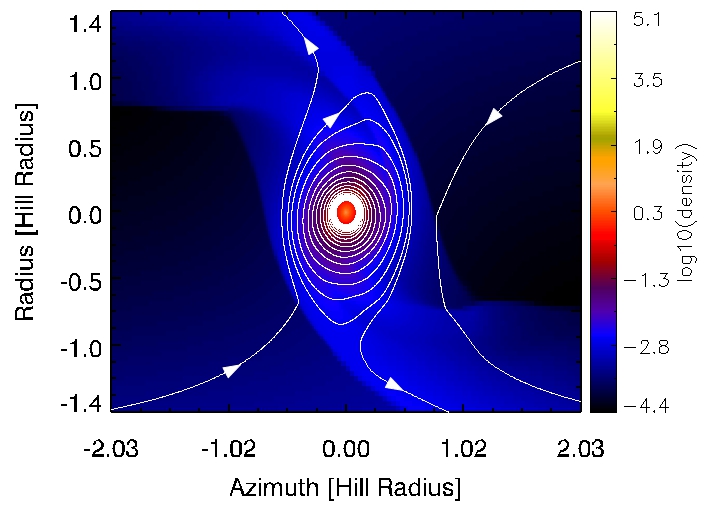} 
\caption{Volume density map on the midplane in the vicinity of a Jupiter-mass planet for our viscous, low-resolution simulation. A few streamlines are also shown with arrows showing the direction of the flow.}
\label{stream-visc}
\end{figure}

We find that the direction of the radial flow on the midplane of the CPD depends strongly on viscosity. In our viscous simulation the flow is outwards, as shown by the streamlines plotted in Fig.~\ref{stream-visc}. The outflow near the midplane -- together with inflow in the upper layers -- is indeed typical of a three dimensional viscous-accretion disk (see \citealt{Urpin84, S88, KL92, R94, RG02, TL02}).

\begin{figure}
\includegraphics[width=\columnwidth]{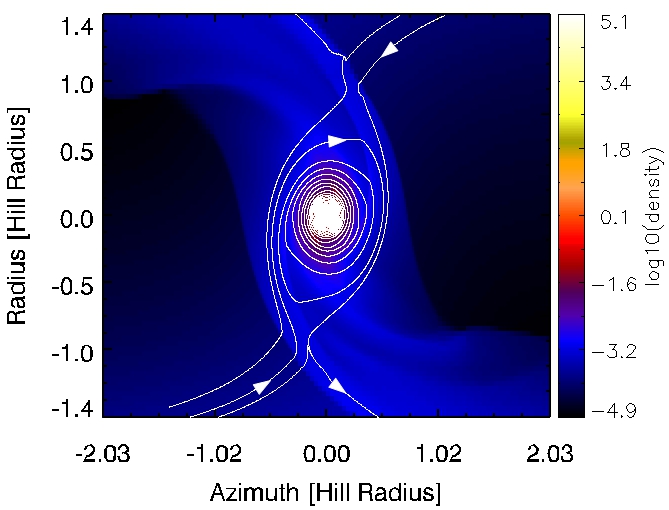} 
\caption{Same as Fig. \ref{stream-visc}, but for the nominal (i.e. inviscid) simulation. One of the streamlines shows clear shocks when crossing the spiral density wave, thus the gas flow loses angular momentum and spirals down to the planet. The planet accretes all the gas flowing between the first and the third streamlines from the bottom left of the figure and between the first and the second streamlines from the top right of the panel.}
\label{streamline}
\end{figure}

In our nominal simulation, instead, the net flow is inwards. This is due to two reasons: (I) the effective viscosity is smaller and (II) the flow suffers more pronounced shocks when crossing the spiral density wave. The latter issue is well visible in Fig.~\ref{streamline}. The shocks correspond to the points where the streamlines change abruptly direction. Look in particular at the accreting streamline on the figure. When it encounters the wave for the first time, the streamline changes abruptly direction relative to the position of the planet. The streamline now makes a hyperbolic arc around the planet. If unperturbed, it would leave the planet's sphere of influence, but it is shocked again when crossing the outer branch of the density wave at the apex of its trajectory. The shock deviates the motion once again, and reduces its angular momentum relative to the planet.  The streamline now makes a downwards arc around the planet with a large eccentricity, but then it is shocked again and again, every half orbit. Each shock causes a loss in angular momentum, so that the streamline spirals towards the planet.

The shocks were also visible in Fig.~\ref{stream-visc}, but they were less pronounced. The net flow is the result of the competition between the viscous stress, which pushes the flow outwards, and the shocks, which cause angular momentum losses. In the viscous simulation the former wins; in our nominal simulation the latter win. The same competition should occur also for CPDs with adiabatic EOS. Shocks are weaker in that case \citep{DAngelo03}, but in the limit of zero viscosity they should dominate nevertheless.

It is unclear to us why \citet{Tanigawa12} found outflow in their simulation, which had no prescribed viscosity as in our nominal case. Possibly the numerical viscosity in their simulation was higher than in ours; or, alternatively, the fact that no gap opened in their simulation changed substantially the local dynamics.

We stress, however, that the discussion about the direction of the midplane flow in the CPD is mostly academic. In fact, even in the case of inflow on the midplane, the delivery of material to the CPD along the midplane is only 10\% of that due to the vertical inflow. We derived this percentage through the following procedure. First, we plotted the azimuthally averaged mass-flux on circles at different planetocentric radii. The mass-flux is increasing with decreasing distance because mass is continuously added to the CPD from the vertical direction. Considering a distance of 0.58 Hill radii from the planet, which corresponds to the largest radius at which all streamlines are accreting (see Fig. \ref{streamline}) the mass-flux is 10\% of the planet's accretion rate (see ``mass-cut" accretion rate in Section \ref{accrete} for details). Thus,  the remaining 90\% of the accretion has to come from the vertical direction.

\section{Planetary Accretion}
\label{accrete}

After having analyzed in details the dynamics in the vicinity of the planet, we are now ready to discuss the planet's accretion rate. 

First, we checked whether we reached a stationary state at the end of the simulation by comparing the radial mass fluxes (averaged over azimuth and integrated vertically) obtained at different output times throughout the circumstellar disk and the CPD. Having concluded positively that a quasi stationary state was reached, we then checked whether the flux of mass onto the planet was a simple consequence of the mass flux towards the star in the circumstellar disk. We found, as discussed more in details in \citep{Morbidelli13}, that the flux of gas towards the planet is due to the flux of gas into the gap from both of its sides. Thus, the planet accretion rate would not be zero even in an equilibrium disk without any net mass flux to the star. Our disk is indeed very close to an equilibrium disk for $\alpha=0$\,; the flux of gas towards the star in our disk is not significant, and therefore not correlated to the accretion of the planet.

In order to measure the accretion rate in the simulation we measured how much mass was removed through the mass-cut. In our nominal simulation after reaching a stationary state, we found a large accretion rate, namely $\dot{M} = 2 \times 10^{-7} \rm{M_*} \Omega$. Again, the results scale linearly with $\Sigma_0$ and the relationship derived in Section \ref{setup} is $\Sigma_0=\Sigma_{MMSN} \sqrt{a_{jup}/a} (M_*/M_{\odot})$. Moreover, $\Omega=\sqrt{GM_*/a^3}=\sqrt{\frac{M_*}{M_{\odot}}\frac{GM_{\odot}}{(1\rm{AU})^3}\frac{(1\rm{AU})^3}{a^3}} $. Plugging in these will lead to $\dot{M} = 5.51 \times 10^{-7}\rm{M_{\odot}}/\rm{year} \times (M_*/M_{\odot})^{1/2} \times (\Sigma_0/\Sigma_{MMSN}) \times (a/1\rm{AU})^{-1}$. If the planet is at 5.2 AU, this corresponds to $\sim 10^{-4}$ Jupiter masses per year. We argue that this high accretion rate is due to numerical viscosity. In fact, in the high resolution simulation, where the numerical viscosity is halved, the accretion rate is reduced by a factor of two. 

Interestingly, in the high resolution simulation, the mass in the CPD is basically the same as in the nominal simulation (see Fig.~\ref{coldens}). This is because the polar inflow is also reduced by a factor of two. This is at first surprising, because in 2D disks at low viscosity the width and depth of a gap is independent of viscosity \citep{Crida}. But 3D gaps behave differently. The detailed analysis of the gas dynamics in a 3D gap will be the object of another paper \citep{Morbidelli13}.  But in brief, the dynamics of a gap in a 3D disk is characterized by an interesting circulation\,: the gas flows into the gap from the surface of the circumstellar disk, then precipitates towards the midplane. In doing this, it falls either to the CPD or gets kicked out by the planet and
goes back into the circumstellar disk. The flow into the gap at the disk's surface is dominated by the numerical viscosity and so it changes by a factor of two from the nominal to the high resolution simulation. 

This result shows that it is not possible to assess the accretion rate of a planet in the low-viscosity limit just using simulations with no prescribed viscosity. This has to be kept in mind regardless of the EOS used in the simulations. Instead, we need to identify and quantify the accretion mechanisms that are independent of viscosity. However, we remind that our analysis is based on isothermal simulations; if a more realistic EOS is used, the quantitative relevance of each mechanisms may change. The analysis below, therefore, should be regarded as a proof of concept, useful also for future radiative studies, and not for its quantitative results (see also Sect.~\ref{discussion}).

Here we envision a scenario in which the circumstellar disk has a layered structure, with a dead zone near the midplane and an active viscous layer near the surface, in agreement with MRI studies \citep{Gammie96}. We envision also that CPD is mostly MRI-inactive, in agreement with  \citep{Fujii11,Turner13,Fujii13}, so we investigate the planet accretion rate in the limit of vanishing viscosity in the CPD.

A first mechanism of accretion, independent of the viscosity in the CPD, is the vertical inflow. We stress that the vertical inflow is sustained by the flux of gas in the active layer of the circumstellar disk \citep{Morbidelli13}, so that it should exist also if the planet forms in a dead zone and the CPD is inviscid. We have seen in Section ~\ref{vertical} that the vertical inflow has a specific angular momentum significantly smaller than the CPD.  Because of the contact of the inflow and the CPD happens through a shock, the inflow subtracts angular momentum from the CPD even in the limit of zero viscosity. Nevertheless, if the specific angular momentum of the incoming gas is larger than that corresponding to an orbit at the surface of the planet, the inflow cannot promote accretion onto the planet. Therefore, the mass accreted by the planet cannot be larger than the mass carried by the inflow of gas with a specific angular momentum smaller than ($j<\sqrt{GM_pR_p}$) \citep{Tanigawa12}. We call ``polar inflow" this subset of the vertical inflow.  

To estimate the accretion rate due to the polar inflow we proceed as follows. We set the radius of the planet to be equal to twice the current radius of Jupiter. This is because the planet at the accretion time was much hotter and therefore its radius was inflated by more or less a factor of two \citep{Jupiter-book}. Also, we refer to the viscous simulation. The reason is that, as we said above, the vertical infall is fed by the gas entering into the gap at the surface of the circumstellar disk, and the latter should be MRI active. With these settings we find an accretion rate of $4 \times 10^{-9} \times M_* \Omega=11.02 \times 10^{-9} \rm{M_{\odot}}/\rm{year} \times (M_*/M_{\odot})^{1/2} \times (\Sigma_0/\Sigma_{MMSN}) \times (a/1\rm{AU})^{-1} $, i.e. $2\times 10^{-6}$ Jupiter masses/yr with the usual scalings and it should scale linearly with $\alpha$. This estimate is one order of magnitude smaller than in \citet{Tanigawa12}, presumably due to the fact that in our simulations the planet opened a gap.

The second accretion process that does not depend on the viscosity in the CPD is the loss of angular momentum in the CPD due to the torque exerted by the star through the spiral density wave \citep*{ML11,Rivier12}. This torque was already considered in \citet{Rivier12} in their 2D simulations. The authors there assumed that in the inviscid case the torque is deposited only in the very inner part of the CPD. However, as we have seen in section 3.3, the wave shocks and removes angular momentum also in the outer part of the CPD. The fact that the wave does not seem to shock in the inner part of the CPD is probably an artifact of numerical viscosity, which increases approaching the planet and smears out the density contrasts, consequently erasing the wave and its shock front. Because the simulation does not allow us to resolve where in the CPD the torque is deposited, in order to provide an upper bound of the planet's accretion rate promoted by the stellar torque we adopt the following simple recipe. We integrate the stellar torque from the planet to the radius where it becomes positive, which is basically at the edge of the CPD; then we estimate the fraction of the CPD mass accreted per unit time as the fraction between the integrated stellar torque and the total angular momentum in the disk.

For both the nominal and the high resolution simulation we derive that the stellar torque promotes the accretion of $3\times 10^{-3}$ of the mass of the CPD per planet's orbital period, i.e. $2.5\times 10^{-4}$ of the CPD mass per year, if the planet is at 5.2 AU around a solar-mass star. The fact that this result is independent on numerical resolution makes us confident of its robustness.

The mass of the CPD in our simulation is only $4\times 10^{-4} \rm{M_J}$. However, if the disk could not accrete onto the planet as fast as in our simulation due to the lack of viscosity, the gas would pile up into the disk, increasing the CPD mass. How massive the disk can become cannot be studied using isothermal simulations and will be the object of a future study. In \citet{Rivier12} it was estimated analytically that the maximum mass of the CPD is $\sim10^{-3} \rm{M_J}$; at this mass its vertical pressure gradient becomes large enough to stop the vertical inflow, so that the mass of the CPD can not grow further. This estimate is probably valid only at the order of magnitude level. However, even assuming a CPD mass of $0.01 \rm{M_J}$, the stellar torque would imply an accretion rate of only $2.5\times10^{-6} \rm{M_J}/\rm{yr}$, i.e. a mass doubling time of 400,000 years. This timescale is comparable to that of the photoevaporation of the circumstellar disk. If this result is confirmed in future, more realistic studies, it implies that, if giant planets form towards the end of the disk's lifetime, the competition between the planet's accretion timescale and the disk removal timescale might explain the wide range of masses observed for giant planets.

\section{Discussion \& Perspective}
\label{discussion}

In this paper the simulations were all isothermal. 
Previous studies showed that for small planets ($\sim 10 \,\rm{M_{\oplus}}$) the flow near the planet strongly depends on the equation-of-state (\citealt{PM08}, \citealt{AB09}, \citealt{NR13}), but for Jupiter-mass planets the accretion rate in non-isothermal simulations is close to that in isothermal calculations \citep{Machida10}. 

However, we suspect that the \citet{Machida10} result is due to the large planet's accretion rate, consequence of numerical viscosity. At the level of detail at which we explored the local dynamics in this paper, we expect that the equation of state would strongly influence the results at a quantitative level. In particular, in the limit of vanishing viscosity, the gas should pile up in the CPD, and an adiabatic equation of state, with flux limited energy transfer is expected to change significantly the final equilibrium structure of the CPD relative to the isothermal equation-of-state. 

The issue of the pile-up of material in the CPD is crucial to estimate the planet's accretion rate in the inviscid limit. If the mass of the CPD becomes large, the stellar torque can be sufficient to promote a fast accretion onto the planet. \citet*{ML12AAS} suggested that the disk may become gravitationally unstable, which would cause FU Orionis-like accretion bursts onto the planet. In fact, works comparing the results in isothermal and radiative simulations, such as \citet{DAngelo03} or \citet{Gressel13}, show that the CPD tends to be {\it less massive} if adiabatic heating and radiative effects are taken into account. However these works are affected by a large viscosity -- prescribed or numeric -- which prevents the pile-up of mass in the CPD. It should be investigated what actually happens in the ideal inviscid limit.

Nevertheless, it may be possible that the CPD becomes so hot and vertically extended that it does not allow the accretion of new material from the vertical inflow. 
The velocity of the vertical inflow could become sub-sonic; there would be no shock at the surface of the CPD and the flow could be diverted by the pressure gradient. Indeed, radiative simulations like \citet{AB09} and \citet{Gressel13} show that the boundary between the CPD and the vertical flow is less sharp than in our study, suggesting a weakening of the shock front. If the vertical inflow is diverted before that the CPD becomes gravitationally unstable, then a steady state equilibrium can be reached. Like in Sect.~\ref{accrete} the accretion rate onto the planet will depend on the mass of this steady state CPD and the stellar torque, but the quantitative estimate will presumably be different from the one achieved in this paper. We also notice that radiative simulations (e.g. \citealt{Gressel13}) show that the spiral wave launched by the star in the CPD is much less prominent than in the isothermal case, which would reduce the stellar torque.

In the future it will be necessary to investigate how the results change from the quantitative point of view if a more realistic EOS is used. To study the pile-up in the CPD, though, one will still have the problem of numerical viscosity. If the latter promotes the accretion of material from the CPD to the planet, the final mass distribution in the CPD will not be the same as in the ideal inviscid case. Particular care will be needed to address this issue.

\section{Conclusions}
\label{conclusion}

In this paper we studied the dynamics of gas in the vicinity of a Jupiter-mass planet and the properties of the circumplanetary disk. For this purpose we used the {\sc Jupiter} code, a 3D nested-grid hydrodynamical code. We performed locally isothermal simulations with two prescribed $\alpha$ viscosities ($\alpha=0.004$ and $\alpha=0$) and, for $\alpha=0$, with two different resolutions. 

Our results confirm those of \citet{AB09,Machida10,Tanigawa12} concerning the vertical inflow and the CPD vertical structure. We have pointed out, however, that the CPD upper layer is wavy, i.e. the aspect ratio of the CPD changes with planetocentric azimuth, due to the pressure of the inhomogeneous vertical inflow. In a reference frame rotating with the gas around the planet (at a given radius), this pressure exerts a periodic perturbation, leading to the formation of a stationary wave in the CPD vertical structure.  We also were able to reduce the viscosity more than previous local box simulations; in our inviscid simulation the shocks were more pronounced.

We found that CPD is mostly sub-Keplerian, similarly to \citet{Tanigawa12}, and \citet{Uribe13}, except in its upper layer, where it can be slightly super-Keplerian due to the significant flaring of the disk. The radial extent of the disk where the streamlines are quasi-circular depends on viscosity and, if $\alpha=0$, also on numerical resolution. The smaller is the effective viscosity, the smaller is the circular portion of the disk. 

We found that the flow in the CPD midplane is inwards if $\alpha=0$, in contrast with \citet{Tanigawa12}, and \citet{AB12}. In this case the gas flow in the CPD is crossing the spiral density wave twice in every orbit, and each crossing leads to the loss of angular momentum due to a shock. Thus the flow spirals down to the planet. Nevertheless, we showed that the radial inflow of mass through the outer boundary of the CPD is only 10\% of the gas influx coming from the vertical direction. Instead, in the case of the viscous simulation with $\alpha = 0.004$, the flow is spiraling outwards in the midplane. Therefore one can conclude, that the viscosity determines the directions of the flow in the CPD. 

Our simulation resulted into a high planetary accretion rate, namely $\dot{M} = 1\times 10^{-4}$ Jupiter masses per year for a Jupiter mass planet at 5.2 AU in a MMSN; however we showed that this high rate is due to numerical viscosity. We identified that the main accretion mechanisms, independent of  viscosity, is  the torque exerted by the star onto the CPD. We found that the stellar torque promotes the accretion of $2.5\times 10^{-4}$ of the mass of the CPD per year, assuming a planet's orbital period of 12 years. However, we cannot provide a reliable estimate of the mass of the CPD with our isothermal simulations, particularly in the limit of vanishing viscosity, which could lead to a significant pile-up of material in the CPD. An order of magnitude analytic estimate in \citet{Rivier12} reported a CPD mass of $\sim 10^{-3} \rm{M_{J}}$. 
Even assuming a CPD mass of $0.01 \rm{M_{J}}$, the stellar torque would lead to an accretion rate of only $2.5\times10^{-6} \rm{M_{J}}/\rm{yr}$. In other words a Jupiter would build up in 400,000 years with this accretion rate. This timescale is comparable to the removal timescale of the circumstellar disk gas \citep[e.g.][]{Koepferl13,Gorti09,Szulagyi12}.

Although future simulations implementing a realistic, non-isothermal equation of state are needed to achieve a reliable quantitative estimate of the planet's accretion rate in the limit of vanishing viscosity, many conceptual results of this paper, particularly those on the role of numerical viscosity and viscosity-independent transport mechanisms in the CPD, should be valid also in a more realistic context.

The main result presented in this paper is encouraging. The similarity between planet accretion and disk removal timescales suggests that, if the giant planets form towards the end of the disk's lifetime, the competition between the planet's accretion process and disk's photoevaporation could explain the observed, wide range of giant planet masses.

\acknowledgments

This work is part of the MOJO project, supported by the French ANR. We are grateful to T. Tanigawa, and B. Bitsch for helpful discussions. We thank to S. Jacobson for the language corrections. J. Szul\'agyi acknowledges the support from the Capital Fund Management's J.P. Aguilar Grant. A. Morbidelli also thanks the German Helmholtz Alliance for providing support through its Planetary Evolution and Life program. F. Masset acknowledges UNAM support through the grant PAPIIT IA101113.

\appendix

\section{Appendix material}
\label{appendix}

Here we present a toy-model for the wavy structure of the CPD. Our goal is to schematize with some basic physical considerations the CPD's reaction to the pressure of the vertical inflow, which is leading to this wavy disk-surface. This toy-model might help to understand better the process, without giving an exact, i.e. complex physical model which is not goal of this paper.

In the reference frame rotating with the fluid elements in the CPD, the periodic excitation by the vertical inflow's pressure creates a stationary wave in the vertical structure of the CPD. The solution for the acoustic wave equation 
\begin{equation}
\frac{1}{c_s^2}\frac{\partial^2p}{\partial t^2}=\frac{\partial^2p}{\partial z^2}
\end{equation}
for a stationary wave can be written in the following form\,:
\begin{equation} \label{eq:stwave}
p(z,t)=2p_0 e^{i\nu t}\cos(kz)
\end{equation}
where $t$ stands for the time, $z$ represents the vertical coordinate, $\nu$ indicates the wave frequency, and $k$ means the wave-number. The term $\cos(kz)$ does not involve a phase because the CPD is supposed to be symmetric with respect to the midplane, so that we have $\partial p/\partial z=0$ at $z=0$. Putting this equation back into the wave equation we get that $k=\nu/c_s=2\pi /\lambda$ where $c_s$ indicates the sound speed and $\lambda$ is the wavelength. Because the pressure exerted by the vertical inflow has a frequency that is twice the planetocentric orbital frequency ($\nu=2\omega$), then at $z_{max}$ the equation can be written as $p(z_{max},t)=p_{z_{max}} e^{i2\omega t}$. If we equal this with Equation \ref{eq:stwave}, then we get $\lambda=\pi c_s/ \omega=\pi H_{CPD}$.

This is precisely what is seen in Fig.~\ref{disk_surface}. The figure shows the vertical profile of the volume density in the CPD, for various values of the azimuth, at a distance $d \approx 0.058$ Hill-radii from the planet, where $H_{CPD}/d \,\approx\, 0.1$. The $z$ coordinates is normalized by $H_{CPD}$. The profiles oscillate around the well-known Gaussian hydrostatic equilibrium profile. One can see two knots, where all curves intersect, corresponding to the locations in $z$ where the amplitude of the wave is zero, namely corresponding to $\cos(kz)=0$\,: $z=(\pi/4)H_{CPD}$ and $z=(3\pi/4)H_{CPD}$. The distance between the two knots is $\lambda/2=\pi/2H_{CPD}$.
The computation of the cumulated mass along the curves shown in Fig.~\ref{disk_surface} reveals that $>97\%$ of the disk mass is below the knot at $\frac{3}{4}\lambda=\frac{3}{4} \pi H_{CPD} \,\approx\, 2.4 H_{CPD}$. Thus, the extreme ``waviness" of the surface of the disk observed in Fig.~\ref{wavy} concerns solely an ``atmosphere" of the disk accounting only for $<3\%$ of the disk mass.

\end{document}